\documentclass[%
  reprint,
  preprintnumbers,
  superscriptaddress,
  amsmath, amssymb,
  aps,
  prc,
  floatfix,
  nolongbibliography,
]{revtex4-2}

\usepackage{graphicx}
\usepackage{color}
\usepackage[breaklinks,colorlinks,urlcolor=blue,citecolor=blue,linkcolor=blue]{hyperref}
\usepackage{palatino}
\usepackage{mathpazo}
\usepackage{dcolumn}
\usepackage{booktabs,multirow,makecell}
\usepackage{tikz}
\usetikzlibrary{arrows,chains,matrix,positioning,scopes,shapes}

\graphicspath{{figures/}}

\newcommand{\TThrust}{\ifmmode{\tau_{\perp}}%
  \else
  \mbox{$\tau_{\perp}$}
  \fi %
}
\newcommand{\Btot}{\ifmmode{B_{tot}}%
  \else
  \mbox{$B_{tot}$}
  \fi %
}

\newcommand{\snn}[2]{\ifmmode{\sqrt{s}=#1~\mathrm{#2}\ignorespaces}%
  \else
  \mbox{$\sqrt{s_{\mathrm{NN}}}=#1~\mathrm{#2}$}\ignorespaces
  \fi
}
\newcommand{\POWHEG}{\textsc{Powheg}}
\newcommand{\PYTHIAe}{\textsc{Pythia8}}
\newcommand{\FASTJET}{\textsc{FastJet}}
\newcommand{\GeV}{\ifmmode{\mathrm{~GeV}}%
  \else
  {~GeV}
  \fi
}
\newcommand{\pp}{\ifmmode{pp}%
  \else
  \mbox{$pp$}\ignorespaces
  \fi
}
\newcommand{\pbpb}{\ifmmode{\mathrm{PbPb}}%
  \else
  \mbox{PbPb}\ignorespaces
  \fi
}
\newcommand{\njet}{\ifmmode{N_{\mathrm{jet}}}\ignorespaces%
  \else
  \mbox{$N_{\mathrm{jet}}$}\ignorespaces
  \fi
}
\newcommand{\lnbtot}{\ln(B_{tot})}

\begin{document}

\title{Multijet topology in high-energy nuclear collisions: jet broadening}

\author{Jin-Wen Kang}
\affiliation{
  Key Laboratory of Quark \& Lepton Physics (MOE) and Institute of Particle Physics,
  Central China Normal University, Wuhan 430079, China
}

\author{Lei Wang}
\affiliation{
  Key Laboratory of Quark \& Lepton Physics (MOE) and Institute of Particle Physics,
  Central China Normal University, Wuhan 430079, China
}%

\author{Wei Dai}%
\affiliation{%
  School of Mathematics and Physics, China University of Geosciences, Wuhan 430074, China
}%

\author{Sa Wang}%
\affiliation{
  Guangdong Provincial Key Laboratory of Nuclear Science, Institute of Quantum Matter,
  South China Normal University, Guangzhou 510006, China
}

\author{Ben-Wei Zhang}
\email{bwzhang@mail.ccnu.edu.cn}
\affiliation{
  Key Laboratory of Quark \& Lepton Physics (MOE) and Institute of Particle Physics,
  Central China Normal University, Wuhan 430079, China
}%

\date{\today}

\begin{abstract}
  This work presents the first theoretical investigation of the medium modification of jet broadening as an event-shape observable in multijet final states due to jet quenching in high-energy nuclear collisions. The partonic spectrum of \pp\ collisions with next-to-leading order (NLO) accuracy at $\sqrt{s_{\mathrm{NN}}} = 5.02$ TeV is provided by the \POWHEG+\PYTHIAe\ event generator, and the linear Boltzmann transport (LBT) model is utilized to investigate the energy loss of fast partons as they traverse through the hot and dense QCD medium. Jet broadening distributions in multijet final states for both \pp\ and \pbpb\ collisions at $\sqrt{s_{\mathrm{NN}}} = 5.02$ TeV are calculated. We observe an enhancement at the small jet broadening region while a suppression at the large jet broadening region in \pbpb\ collisions relative to that in \pp. This suggests that medium modifications with parton energy loss in the QGP lead to a more concentrated energy flow in all observed multijet events in \pbpb\ reactions.
  We also demonstrate that the intertwining of two effects, the jet number reduction and the restructured contribution, results in the novel behavior of nuclear modification of the jet broadening observable in \pbpb\ collisions.
\end{abstract}

\maketitle

\section{Introduction}

In heavy-ion collisions (HICs) at the RHIC and the LHC, a new form of nuclear matter,  the quark-gluon plasma (QGP)~\cite{Busza:2018rrf,Yagi:2005yb,Zhang:2021xib} composed of de-confined quarks and gluons, may be created, and the jet quenching (or parton energy loss effect) has long been proposed as an excellent hard probe to unravel the fascinating properties of the QCD medium at extreme high density and temperature~\cite{Bjorken:1982tu,Gyulassy:2003mc,Wang:1992qdg,Qin:2015srf,STAR:2009ojv,CMS:2011iwn,Zhang:2003wk,Xie:2024xbn}.
It has been extensively studied that the parton energy loss effect may result in variations of leading hadron productions such as the suppression of inclusive hadron spectra, the disappearance of away-side di-hadron correlations momentum and other hadron correlations~\cite{Vitev:2002pf,Vitev:2005yg,Zhang:2007ja,Zhang:2009rn,Xu:2010du,Eyyubova:2014dha,Dai:2015dxa,Liu:2015vna,Dai:2017piq,Cao:2017hhk,Albacete:2018ruq,Ma:2018swx,Pandey:2022yiy}.
As the running of heavy-ion programs at the LHC with its unprecedented colliding energies, the study of parton energy loss has been extended to medium modifications of many jet observables~\cite{Bass:2008rv,Vitev:2008rz,Vitev:2009rd,Majumder:2010qh,He:2011pd,Apolinario:2012cg,JET:2013cls,Senzel:2013dta,Blaizot:2015lma,Casalderrey-Solana:2015vaa,Milhano:2015mng,CMS:2017ehl,Connors:2017ptx,Tachibana:2017syd,Dai:2018mhw,Zhang:2018kjl,ATLAS:2019dsv,Wang:2019xey,Cao:2020wlm,ALargeIonColliderExperiment:2021mqf,Cunqueiro:2021wls,Wang:2021jgm,Chen:2022kic,JETSCAPE:2022jer,Liu:2022kzv,Wang:2022yrp,Wang:2023eer,Li:2024uzk}.
However, so far most studies of jets in HICs have focused on inclusive jets, dijet/$Z+\mathrm{jet}$/$\gamma+\mathrm{jet}$ observables, or their substructures, while only very limited attention has been paid to the global geometrical properties of multijet events~\cite{Chen:2020pfa,Mallick:2020dzv}. It is of interest to further explore how parton energy loss influences the global geometry of multijet events and investigate their nuclear modifications with the existence of the QGP.

The geometrical properties of the energy flow of the collisions are usually described by event-shape variables. The event-shape variable is a general term for a large class of observables, including the thrust, the jet broadening, the jet mass, the third-jet resolution parameter, the sphericity, the $\mathcal{F}$-parameter, the 1-jettiness, the Fox-Wolfram moments, and the aplanarity, etc~\cite{Sjostrand:2006za,Banfi:2010xy,ATLAS:2012tch,CMS:2014tkl,ATLAS:2016hjr,ATLAS:2020vup,Hessler:2021usr}.
These event-shape variables are not only used to extract the strong coupling $\alpha_s$ from properties of the final-state~\cite{CDF:1991etg,Catani:1992jc,DELPHI:1996oqw,Biebel:1999zt,DELPHI:2003yqh,Wang:2021tak}, but also to tune/test the parton showers and non-perturbative components of Monte Carlo event generators~\cite{DELPHI:1996sen,Sjostrand:2006za,Kundu:2019scu,Ghosh:2022zdz}.
Previous measurements of event-shape variables have been performed in electron-positron, deep inelastic scattering (DIS), or proton-antiproton collisions. Recently, some experimenters have focused on the event-shape variables at large transverse momentum transfer in \pp\ collisions~\cite{ATLAS:2012tch,CMS:2014tkl,ATLAS:2016hjr,ATLAS:2020vup}. Meanwhile, some theorists have begun to explore the event-shape variables in heavy-ion collisions at the LHC energies. In Refs.~\cite{Chen:2020pfa,Mallick:2020ium,Mallick:2020dzv,Prasad:2021bdq}, the authors studied the transverse sphericity in heavy-ion collisions at the LHC energies. It is noted though for transverse sphericity two-jet events may give a large contribution, for jet broadening as an event-shape observable defined to utilize the transverse thrust axis the contribution of two-jet events turns out to be zero. Therefore, jet broadening should provide relatively more complete information on the geometric properties of the energy flow of multijet events with jet number $\njet \geqslant 3$.

In this paper, we present the first theoretical study on the normalized distribution of jet broadening (denoted as $\Btot$) in \pbpb\ collisions. We employ \POWHEG+\allowbreak\PYTHIAe\ \cite{Nason:2004rx,Frixione:2007vw,Alioli:2010xd,Alioli:2010xa,Sjostrand:2014zea}, a Monte Carlo model matching NLO matrix elements with parton shower, to obtain baseline results of jet broadening distributions in \pp\ collisions. Our model calculations of jet broadening distribution could provide a decent description of CMS data in \pp\ collisions. Then we calculate the medium modification of the jet broadening distribution in multijet ($\njet \geqslant 3$) final-states in \pbpb\ collisions at \snn{5.02}{TeV}\ for the first time. We show the normalized distribution of jet broadening \Btot  is enhanced at small \Btot region in \pbpb\ collisions (while suppressed at larger $\Btot$) as compared to that in \pp\ collisions. It indicates the selected events with lower jet broadening \Btot increase due to the jet quenching effect. In other words, the energy flow patterns of the collision events in \pbpb\ become more concentrated after including parton energy loss effect.

This paper is organized as follows. Sec. \ref{sec:framework} describes details of the theoretical frameworks, including the definition of the observables, the generation of the \pp\ baseline, and the energy loss model. In Sec.~\ref{sec:res}, we present the numerical results of the normalized jet broadening distributions in \pbpb\  collisions, and discuss how jet quenching impacts the jet broadening in heavy-ion collisions. We give a brief summary in Sec.~\ref{sec:summary}, and also include an Appendix, which provides a detailed calculation of the concerned observables in the ideal symmetric multijet limit.

\section{Framework}
\label{sec:framework}

In this work, we consider multijet event-shape observables that measure the properties of the energy flow in the final-states of high-energy particle collisions. Because performing measurements near the beam are complex, one way to address this difficulty is to define event-shapes using only jets in a central region. Here, we utilize those jets in the central rapidity region to calculate the event-shapes, and we select those jets that satisfy $|\eta_{\mathrm{jet}}|<2.4$.

The transverse thrust \TThrust is defined as~\cite{ATLAS:2012tch,CMS:2014tkl,ATLAS:2020vup}
\begin{equation}
  \TThrust \equiv 1 - \max_{\hat{n}_T} \frac{\sum_i \left|%
    \vec{p}_{T,i}\cdot \hat{n}_T\right|
  }{\sum_i p_{T,i}},
\end{equation}
where $\hat{n}_T$ is the transverse thrust axis and it is the unit vector that maximizes the sum of the projections of $\vec{p}_{T,i}$, $p_{T,i}$ represents the transverse momentum of the $i$-th jet. This variable is sensitive to the modeling of two-jet and multijet ($\geqslant 3~\mathrm{jet}$) topologies. 
In a perfectly balanced back-to-back two-jet event, \TThrust is zero, and in an isotropic multijet event, it closes to $1-2/\pi$~\cite{CMS:2014tkl,Banfi:2010xy} (see more details in the Appendix~\ref{sec:appendix}).
With a transverse thrust axis $\hat{n}_T$, one can separate the region $\mathcal{C}$ into an upper (U) side $\mathcal{C}_U$ consisting of all jets in $\mathcal{C}$ with $\vec{p}_T \cdot \hat{n}_T > 0$ and a lower (L) side $\mathcal{C}_L$ with $\vec{p}_T \cdot \hat{n}_T < 0$.
Here, $\mathcal{C}$ represents the region that satisfies the given constraints in the space.
Given the event region $X$, we can introduce the mean transverse-energy weighted pseudo-rapidity $\eta_{X}$ and azimuthal angle $\phi_X$ of this region~\cite{CMS:2014tkl,ATLAS:2020vup,Banfi:2010xy},
\begin{equation}
  \eta_X \equiv \frac{\sum_{i\in\mathcal{C}_X}p_{T, i}\eta_i}{\sum_{i\in\mathcal{C}_X}p_{T,i}},\quad
  \phi_X \equiv \frac{\sum_{i\in\mathcal{C}_X}p_{T, i}\phi_i}{\sum_{i\in\mathcal{C}_X}p_{T,i}},\label{equ:def:etaphiX}
\end{equation}
where $X$ refers to upper (U) or lower (L) side. Then, the jet broadening variable in two regions
is defined as
\begin{equation}
  B_X \equiv \frac{1}{2P_T}\sum_{i\in\mathcal{C}_X}p_{T,i}
  \sqrt{\left(\eta_i-\eta_X\right)^2+\left(\phi_i-\phi_X\right)^2},
\end{equation}
where $P_T$ is the scalar sum of the transverse momenta of all the jets in the region. From which
one can obtain total jet broadening,
\begin{equation}
  B_{tot} \equiv B_U + B_L.
\end{equation}
Thus $B_{tot}=0$ for a two-jet event by definition.

\begin{figure*}[htbp]
  \centering
  \begin{minipage}{0.32\textwidth}
    \includegraphics[width=\textwidth]{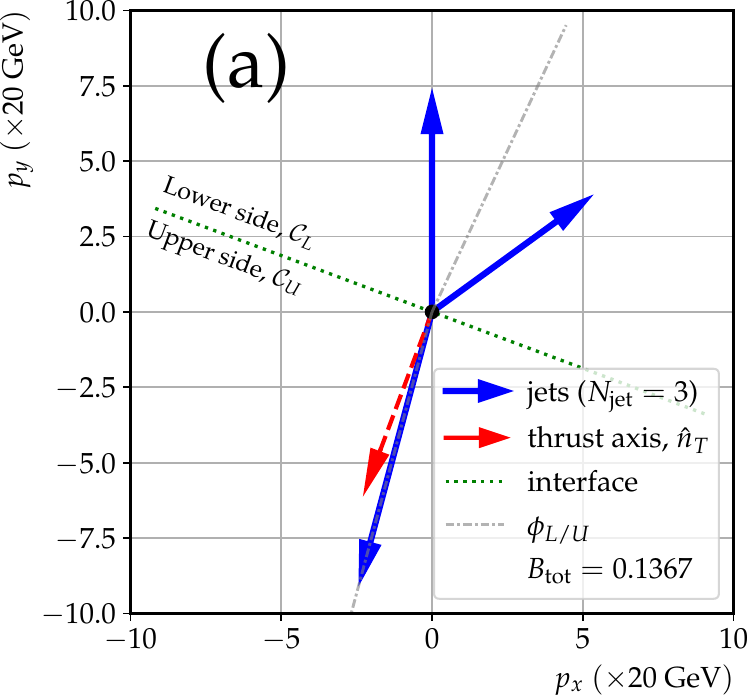}
  \end{minipage}
  \begin{minipage}{0.32\textwidth}
    \includegraphics[width=\textwidth]{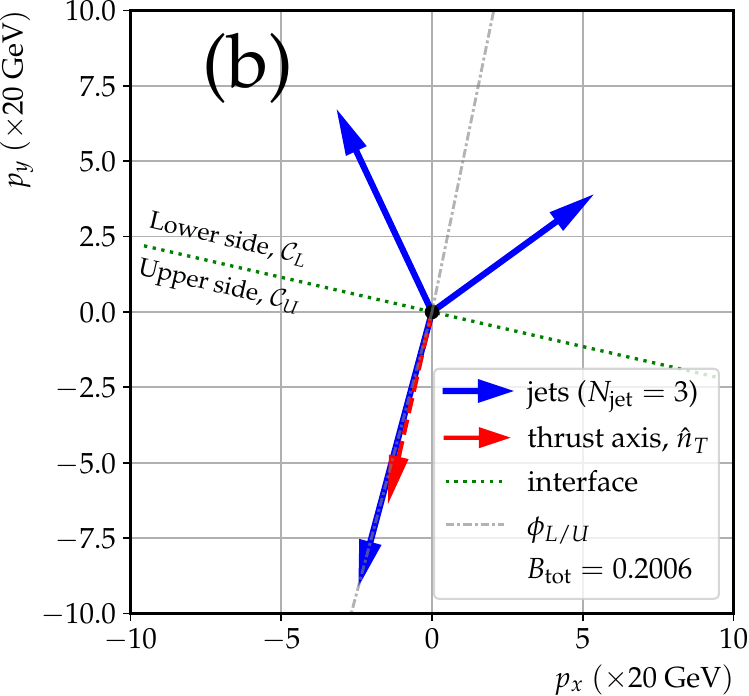}
  \end{minipage}
  \begin{minipage}{0.32\textwidth}
    \includegraphics[width=\textwidth]{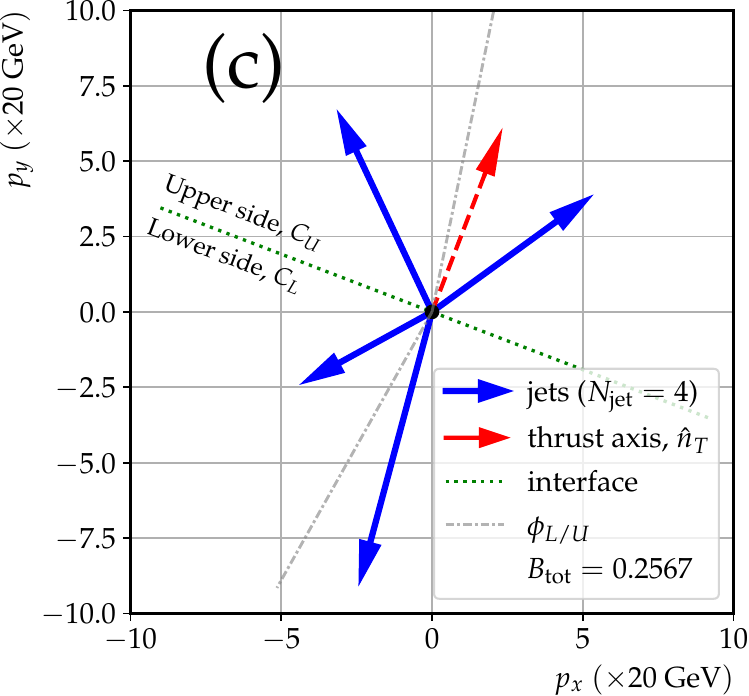}
  \end{minipage}
  \caption{Schematic illustrations of multijet event configuration in the transverse plane and the values of their jet broadening observable.
  The red arrow in the above figures represents the thrust axis, which is originally a unit vector, but we have magnified it to make it easier to see in the plots. It is easy to see that the sign of the $\hat{n}_T$ vector only affects the identification of the upper (or lower) region, and does not change the value of total jet broadening.}
  \label{fig:configurations}
\end{figure*}

We may observe that the descriptions of $\eta_X$ and $\phi_X$ given in Eq.~\eqref{equ:def:etaphiX} are similar to the definition of the center of mass, so we can in some sense regard $(\eta_X, \phi_X)$ as the coordinates of the center of jets energy outflow. Therefore, jet broadening can characterize the weighted broadening of the jets relative to the center of the outgoing energy flow. 
We plot three multijet event configurations in the transverse plane shown in Fig.~\ref{fig:configurations} to demonstrate the physical picture of the jet broadening. In Fig.~\ref{fig:configurations} we suppress the variation of pseudo-rapidity for simplicity by imposing $\eta_i = \eta_X = 0$, in which case $\Btot$ has a maximum value of $\pi/8$ for the circular limit (see the Appendix~\ref{sec:appendix}).
The only difference between the events described in Fig.~\ref{fig:configurations}(b) and Fig.~\ref{fig:configurations}(a) is that the angle between the second and third jets is larger, which makes the total jet broadening of the event configuration shown in Fig.~\ref{fig:configurations}(b) larger than that shown in Fig.~\ref{fig:configurations}(a).
The event in Fig.~\ref{fig:configurations}(c) has one more jet than in Fig.~\ref{fig:configurations}(b), and thus the total jet broadening of the event configuration shown in Fig.~\ref{fig:configurations}(c) is larger than the one in Fig.~\ref{fig:configurations}(b). From Fig.~\ref{fig:configurations}, we can find that the weighted broadening of energy flow can also be used to distinguish the balance of the spatial distribution of energy flow.
Briefly speaking, the multijet energy flow broadening is very small when the jet broadening variable tends to zero; at this point, the spatial distribution of energy flow tends to be very imbalanced. On the contrary, the multijet energy flow broadening increases when the jet broadening variable is away from zero; the spatial distribution of energy flow walks in a more balanced direction.
By the way, we may use the term ``more concentrated'' to express the meaning of ``less broadening'' or ``narrower'' of jet broadening distributions.

A NLO$+$PS Monte Carlo event generator is employed in this research to simulate jet productions in \pp\ collisions. Specifically, the NLO matrix elements for the QCD dijet processes are provided by \POWHEG\ matches with the final-state parton showering in \PYTHIAe\ \cite{Nason:2004rx,Frixione:2007vw,Alioli:2010xd,Alioli:2010xa,Sjostrand:2014zea}. For every event, jets are reconstructed with the anti-$k_t$ algorithm and distance parameter $R=0.5$ using \FASTJET\ \cite{Cacciari:2008gp,Cacciari:2011ma}.
In calculations, we make the kinematic cuts adopted in the CMS publication~\cite{CMS:2014tkl}: the selected events are required to include at least three jets in the central pseudo-rapidity region $|\eta_\mathrm{jet}|<2.4$, the lower threshold of the reconstructed jets $p_T$ is $30\GeV$, and the leading jet $p_T$ satisfies $110 < p_{T,1}<170\GeV$.

\begin{figure}[htbp]
  \centering
  \includegraphics[width=0.95\columnwidth]{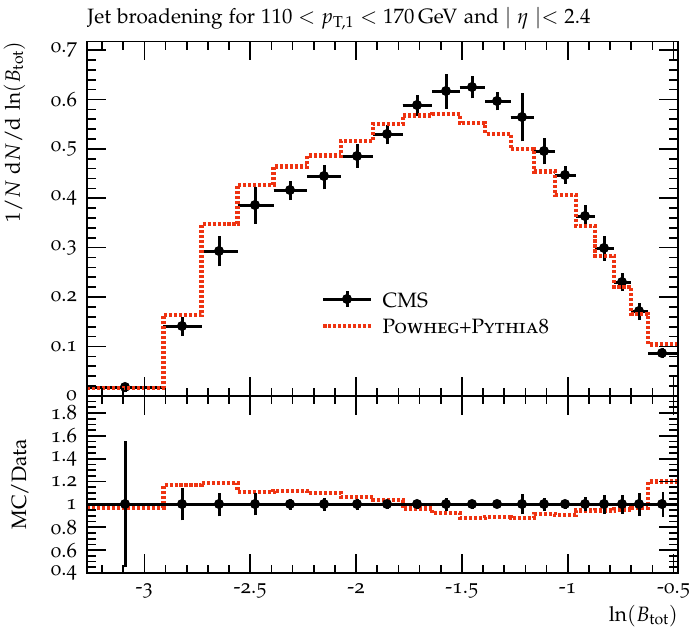}
  \caption{The jet broadening \Btot are calculated at 7~TeV by \POWHEG$+$\PYTHIAe\ in \pp\ collisions and compared to CMS experimental data.}
  \label{fig:ppbaseline}
\end{figure}

Fig.~\ref{fig:ppbaseline} illustrates the multijet event-shape by \POWHEG{}$+$\allowbreak\PYTHIAe\ and the comparison with the CMS data~\cite{CMS:2014tkl} in \pp\ collisions at 7~TeV.
The simulation using \POWHEG$+$\PYTHIAe\ framework can provide a pretty good description of the experimental data in \pp\ collisions. Consequently, it may provide a reliable baseline for further study of medium modifications of event-shape. 
The work~\cite{CMS:2014tkl} of the CMS collaboration points out that the hard processes with two to four final-state partons generated by \textsc{MadGraph}~\cite{Alwall:2011uj,Alwall:2014hca} might make it better match the experimental data. Nevertheless, for technical reasons, we ultimately chose to use \POWHEG\ to produce the \pp\ baseline. 
In our subsequent calculations, we reconstruct the jets at the hadron level to include hadronization corrections.
Furthermore, we used parton samples generated by \POWHEG$+$\PYTHIAe\ in a vacuum as input for the LBT model, ignoring the effects of nuclear parton distribution functions.

In this study, we employ the \textbf{L}inear \textbf{B}oltzmann \textbf{T}ransport (LBT) model to consider both elastic and inelastic scattering processes of the initial jet shower partons and the thermal recoil partons in the QGP medium~\cite{Wang:2013cia,He:2015pra,Cao:2016gvr}. The LBT model has been widely used in the study of jet quenching and successfully describes many experimental measurements~\cite{Chen:2020pfa,Zhang:2018kjl,Luo:2018pto}.

The LBT model uses the linear Boltzmann transport equation to describe the elastic scattering~\cite{Wang:2013cia,He:2015pra,Cao:2016gvr},
\begin{eqnarray}
  p_1\cdot \partial f_1(p_1) &=& -\int\frac{d^3p_2}{(2\pi)^32E_2}
  \int\frac{d^3p_3}{(2\pi)^32E_3}\int\frac{d^3p_4}{(2\pi)^32E_4} \nonumber\\
  &&\times\frac{1}{2}\sum_{2(3,4)}\left(f_1f_2-f_3f_4\right)\left|\mathcal{M}_{12\to 34}\right|^2
  \left(2\pi\right)^4\nonumber\\ &&\times S_2(\hat{s}, \hat{t},
  \hat{u})\delta^{(4)}(p_1+p_2-p_3-p_4),
\end{eqnarray}
where $f_i(p_i)(i=2,4)$ are the parton phase-space distributions in the thermal medium with the local temperature and the fluid velocity; $f_i(p_i)(i=1,3)$ are the phase-space densities for the jet shower parton before and after scattering, the leading-order (LO) elastic scattering matrix elements $\left|\mathcal{M}_{12\to 34}\right|^2$ (Ref.~\cite{Eichten:1984eu} for massless light partons and Ref.~\cite{Combridge:1978kx} for heavy quarks) may be collinear divergent for massless partons, and then are regulated by a Lorentz-invariant regularization condition~\cite{Auvinen:2009qm},
\begin{equation}
  S_2(\hat{s}, \hat{t}, \hat{u})\equiv \theta\left(\hat{s}\geqslant 2\mu_D^2\right)
  \theta\left(-\hat{s}+\mu_D^2\leqslant \hat{t} \leqslant -\mu_D^2\right),
\end{equation}
where $\hat{s}$, $\hat{t}$ and $\hat{u}$ are Mandelstam variables, $\mu_D^2$ is the Debye screening mass.

The inelastic scattering is described by the higher-twist approach~\cite{Guo:2000nz,Zhang:2003yn,Zhang:2003wk,Majumder:2009ge},
\begin{equation}
  \frac{dN_g}{dxdk_{\perp}^2dt}=\frac{2\alpha_sC_AP(x)\hat{q}}{\pi k_{\perp}^4}
  \left(\frac{k_{\perp}^2}{k_{\perp}^2+x^2m^2}\right)^4\sin^2\left(
  \frac{t-t_i}{2\tau_f}\right),
\end{equation}
where $x$ is the energy fraction of the radiated gluon relative to parent parton with mass $m$, $k_\perp$ is the transverse momentum of the radiated gluon, $P(x)$ is the splitting function in vacuum, $\hat{q}$ is the jet transport coefficient~\cite{He:2015pra}, and $\tau_f$ is the formation time of the radiated gluons in QGP medium, with $\tau_f=2Ex(1-x)\big/\left(k_{\perp}^2+x^2m^2\right)$.

The 3$+$1D \textsc{CLVisc} hydrodynamical model~\cite{Pang:2012he,Pang:2014ipa} is employed to generate a dynamically evolving bulk medium with initial conditions from the AMPT model~\cite{Lin:2004en}. Parameters used in the \textsc{CLVisc} are chosen to reproduce hadron spectra with experimental data.

In the LBT model, the strong coupling constant $\alpha_s$ is fixed and is the only parameter that needs to be specified to control the strength of parton interaction. Referring to the published works~\cite{Zhang:2018kjl,Luo:2018pto,Chen:2020pfa}, we take $\alpha_s$ to be 0.2 in this work. To hadronize the partons that have completed their shower evolution within the QGP medium, we employed the minimization criteria introduced in \textsc{JetScape} to construct strings~\cite{Putschke:2019yrg}, and subsequently used \textsc{Pythia}8 to simulate the hadronization and hadron decay processes.

\section{Results and Analysis}
\label{sec:res}

\begin{figure}
  \centering
  \includegraphics[width=0.9\columnwidth]{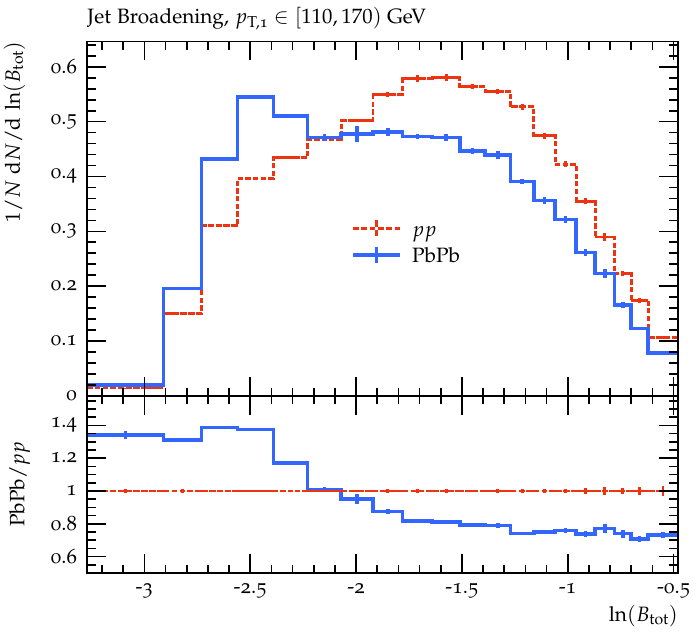}
  \caption{The normalized distributions of the jet broadening \Btot in 0-30\% \pbpb\
    and \pp\ collisions at \snn{5.02}{TeV}. The ratio of the two normalized distributions is given
    in the lower panel. The error bars (vertical lines in the center of the bins) in the figure represent the statistical uncertainties in the data.}
  \label{fig:JB}
\end{figure}

We now present numerical results of jet production in \pbpb\ collisions with jet quenching.
The event selection criteria for both \pp\ and \pbpb\ collisions at $\sqrt{s_{\mathrm{NN}}}=5.02$~TeV are the same, given in Sec.~\ref{sec:framework}.
In order to study the hot nuclear alteration of the jet broadening distribution, we plot Fig.~\ref{fig:JB}, which shows the normalized distributions of the jet broadening $\Btot$ in central 0-30\% \pbpb\ and \pp\ collisions at $\sqrt{s_{\mathrm{NN}}}=5.02$~TeV, and their ratio in the lower panel.
We find wide distributions in \pp\ and \pbpb\ collisions within the range given in the figure.
Most of the contribution of the distribution in \pp\ collisions lies in the region with large $\lnbtot$ values.
Moreover, the normalized distribution in \pbpb\ collisions is shifted toward smaller $\lnbtot$, thus, leading to an enhancement at small $\lnbtot$ region and a suppression at large $\lnbtot$ region. This modification indicates that the proportion of survived events with the narrow jet distribution will increase, after including parton energy loss in the QGP.
Namely, Fig.~\ref{fig:JB} shows that the nuclear modification reduces the multijet energy flow broadening, and thus increases the imbalance of the energy flow.

In addition, we show that the jet broadening distribution in \pbpb\ collisions has a peak at around the $\Btot= 0.082$ ($\lnbtot=-2.5$).
We also notice considerable enhancement at smaller $\lnbtot$ region in \pbpb\ collisions versus \pp\ collisions. Such an enhancement is much more pronounced than some other event-shape observables, such as transverse sphericity.
In the following we will conduct a thorough investigation on the underlying reasons  for these medium modifications of jet broadening observed in Fig.~\ref{fig:JB}.

\begin{figure}[htbp]
  \centering
  \includegraphics[width=0.95\columnwidth]{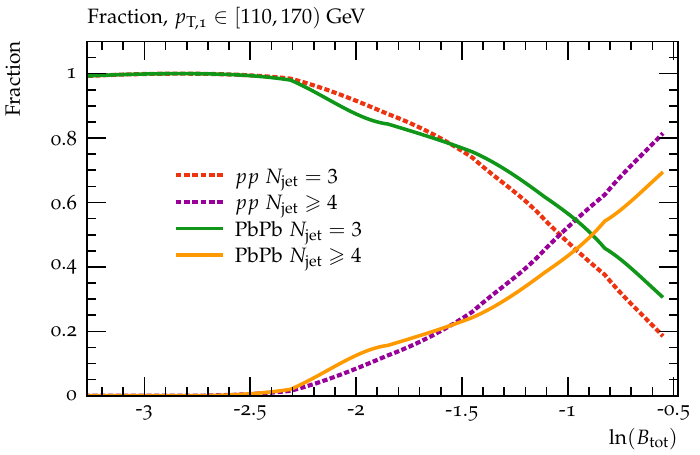}
  \caption{Relative contribution fraction of total event number jet broadening distribution of $\njet=3$, $\njet\geqslant 4$ events in \pp\ and \pbpb\ collisions.}
  \label{fig:relcontrfrac}
\end{figure}

We separate the selected events into two categories: the number of jets in the event equals to three ($\njet=3$) and more than three ($\njet\geqslant 4$).
In Fig.~\ref{fig:relcontrfrac}, we plot the production fractions as functions of $\lnbtot$ for three-jet events and $\njet\geqslant 4$ events respectively in \pp\ and \pbpb\ collisions. We find that in both \pp\ and \pbpb\ collisions, the three-jet events will give dominant contributions at the low $\lnbtot$ region, whereas $\njet\geqslant 4$ events domonate at the large $\lnbtot$ region .

\begin{figure}
  \centering
  \begin{minipage}[c]{0.9\columnwidth}
    \begin{tikzpicture}
      \node[anchor=south west,inner sep=0] (image) at (0,0) {%
        \includegraphics[width=\textwidth]{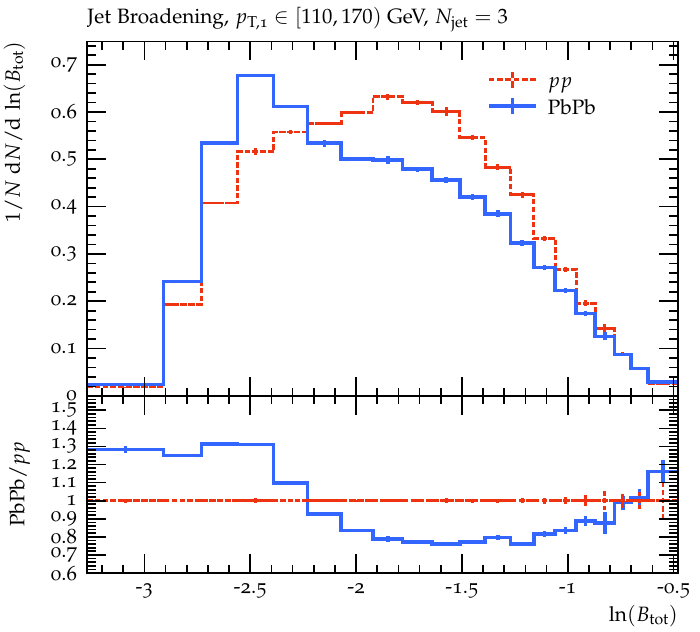}
      };

      \begin{scope}[x={(image.south east)},y={(image.north west)}]
        \node[draw, ultra thick, red] at (0.56, 0.5){$N_{\rm jet} = 3$};
      \end{scope}
    \end{tikzpicture}
  \end{minipage}\\
  \begin{minipage}[c]{0.9\columnwidth}
    \begin{tikzpicture}
      \node[anchor=south west,inner sep=0] (image) at (0,0) {%
        \includegraphics[width=\textwidth]{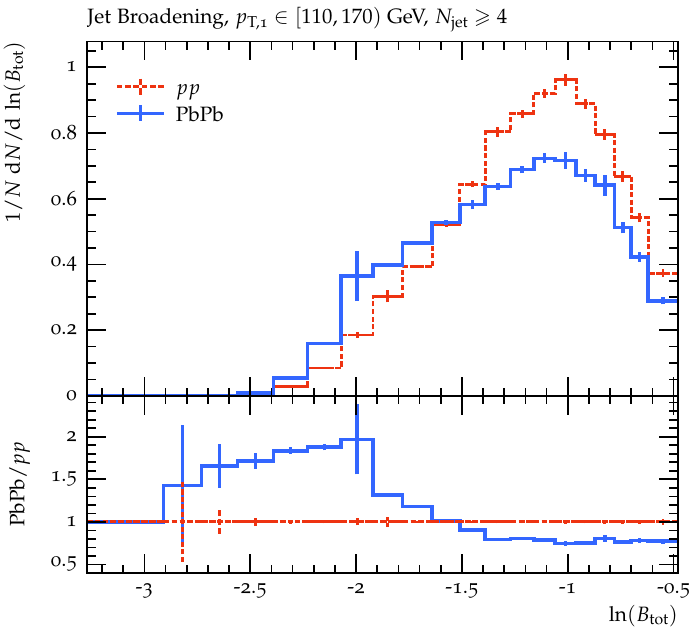}
      };

      \begin{scope}[x={(image.south east)},y={(image.north west)}]
        \node[draw, ultra thick, red] at (0.32, 0.62){$N_{\rm jet} \geqslant 4$};
      \end{scope}
    \end{tikzpicture}
  \end{minipage}
  \caption{Upper (Lower): Normalized jet broadening distributions of $\njet=3$ ($\njet\geqslant 4$) events in \pp\ and \pbpb\
  collisions at \snn{5.02}{TeV}. The distributions are self-normalized. The bottom panels of the two figures give the corresponding medium modification factors.}
  \label{fig:JBDiffJetNum}
\end{figure}

We plot the event normalized $\lnbtot$ distributions of $\njet=3$ events and $\njet \geqslant 4$ events in \pp\ and \pbpb\ collisions at \snn{5.02}{TeV}, as shown in Fig.~\ref{fig:JBDiffJetNum}.
We can find that the $\lnbtot$ distributions of three-jet events and multijet ($\njet\geqslant 4$) events are quite different for both \pp\ and \pbpb\ collisions.
In \pp\ collisions, the $\lnbtot$ distribution of three-jet events has a relatively flat peak near $-1.8$ ($\Btot= 0.165$), while the $\lnbtot$ distribution of $\njet\geqslant 4$ events has a sharp peak near $-1.0$ ($\Btot= 0.368$). For \pbpb\ collisions, the peak of three-jet events shifts to the left near $-2.5$ ($\Btot= 0.082$), which is quite different from the distribution of \pp\ collisions. The peak of $\njet\geqslant 4$ events is still around $-1.0$ compared with that in \pp\ collisions, but it is flatter.
In a word, the distribution of jet broadening for most three-jet events is more concertrated than that for $\njet\geqslant 4$ events. After jet quenching, these distributions become narrower, whether they are three-jet events or events with more than three jets.

Combining the numerical results  in Fig.~\ref{fig:relcontrfrac} and Fig.~\ref{fig:JBDiffJetNum}, we now see that at large $\Btot$ region, the nuclear modifications of jet broadening distributions are mainly determined by the events with $\njet\geqslant 4$.  Since in \pbpb\ the jet broadening distribution in $\njet\geqslant 4$ events at large $\Btot$ region is suppressed as compared to \pp\ (see the lower panel of Fig.~\ref{fig:JBDiffJetNum}), it is understandable that one observes a reduction of total jet broadening (for all events) at large $\Btot$ region in \pbpb\  as illustrated by Fig.~\ref{fig:JB}. Following the same logic, because in \pbpb\ the jet broadening distribution in $\njet =3$ events at small $\Btot$ region is enhanced relative to \pp\ (see the upper panel of Fig.~\ref{fig:JBDiffJetNum}), we then see the increasing of total jet broadening (for all events) at small $\Btot$ region  in \pbpb\ collisions.

We may see the jet number reduction effect plays an important role for the medium modifications of total jet broadening in \pbpb\ collisions.
Due to the jet quenching effect, some jets will fall below the threshold of event selection after energy loss, which decreases the number of jets in such events, and then leads to the modification of event-shape observables~\cite{Chen:2020pfa}.
In Table.~\mbox{\ref{tab:jetnumberfrac}}, we compare the jet number fraction in \pp\ and \pbpb\ collisions at \snn{5.02}{TeV}. The proportion of three-jet events in \pp\ and \pbpb\ collisions at \snn{5.02}{TeV}\ is 76.98\% and 81.17\%, respectively, and the rest are events with four or more jets. In \pbpb\ collisions, the production fraction of three-jet events increases  and the fraction of $\njet\geqslant 4$ events decreases, which may make jet broadening distributions more imbalanced, because $\njet=3$ events tend to give small $\Btot$ while $\njet\geqslant 4$ events tend to contribute more in large $\Btot$ region.
It is noted that the reduction of $\njet\geqslant 5$ events (shown in Table.~\ref{tab:jetnumberfrac}) also underlies why the $\lnbtot$ distribution of $\njet\geqslant4$ events in \pbpb\ collisions of Fig.~\ref{fig:JBDiffJetNum} (lower) is suppressed at large $\lnbtot$.

\begin{table}
  \caption{The relative production fraction of $\njet=3$ events, $\njet =4$ and $\njet\geqslant 5$ events in \pp\ and \pbpb\ collisions at \snn{5.02}{TeV}.
  The kinematic cuts is consistent with the previous text.}
  \label{tab:jetnumberfrac}
  \renewcommand\arraystretch{1.5}
  \begin{ruledtabular}
    \begin{tabular}{ccc}
      & \pp\  & \pbpb\  \\
      \hline
      $\njet=3$ & 76.98$\pm$0.30\%   & 81.17$\pm$0.39\% \\
      $\njet= 4$         & 18.59$\pm$0.13\%   & 15.53$\pm$0.25\% \\
      $\njet\geqslant 5$ &  4.43$\pm$0.05\%   &  3.30$\pm$0.05\% \\
    \end{tabular}
  \end{ruledtabular}
\end{table}


However, is the jet number reduction effect the only reason for the media modifications shown in Fig.~\ref{fig:JB}? To make further exploration, we next track the origin of events that satisfy the selection criteria.
Due to the energy loss effect, the selected quenched (heavy-ion) events and their unquenched version may come from different kinematics regions.
To expose the influence of these contributions from different origins on jet broadening, we consider pairing selected \pbpb\ events with \pp\ events.
The specific operation is selecting quenched (heavy-ion) events satisfying the above selection condition (given in Sec.~\ref{sec:framework}) and then using the matching procedure to match the unquenched (\pp) event corresponding to each selected quenched event.
When we use the LBT model to perform partons evolution, the information output to the \textsc{HepMC3}~\cite{Buckley:2019xhk} file includes the final-state partons (quenched) list and the input partons (unquenched) list. Therefore, this matching procedure is simple and only needs to extract the partons list of the input part in the selected event (quenched).
A similar selection method has been utilized in~\cite{Zhang:2021sua,Brewer:2021hmh}, and we emphasize that in this work we apply the method in the event level instead of the jet level in Refs.~\cite{Zhang:2021sua,Brewer:2021hmh}.

\begin{table}
  \caption{Classification of the unquenched versions of the selected quenched events.}
  \label{tab:eventcategories}
  \renewcommand\arraystretch{1.5}
  \begin{ruledtabular}
    \begin{tabular}{llc}
      \multicolumn{2}{c}{\bf Matched Condition} & \multirow[c]{2}{*}{\bf Category}          \\
      \cline{1-2}
      \multicolumn{1}{c}{Quenched}                                  & \multicolumn{1}{c}{UnQuenched}                                      & \\
      \midrule[0.8pt]
      \multirow[c]{5}{*}{\makecell*[l]{$p_T^{\mathrm{min~jet}} > 30\GeV$\\ $110<p_{T,1}<170\GeV$\\ $\njet \geqslant 3$}} & \makecell*[l]{$p_T^{\mathrm{min~jet}} > 30\GeV$\\ $110<p_{T,1}<170\GeV$\\ $\njet \geqslant 3$\\ (\textit{same as  Quenched})} & Survival \\
      \cline{2-3}
      & \makecell*[l]{$p_T^{\mathrm{min~jet}} > 30\GeV$   \\ $p_{T,1}>170\GeV$\\ $\njet \geqslant 3$} & Falldown \\
      \cline{2-3}
      & \makecell*[l]{Other contribution} & Restructured \\
    \end{tabular}
  \end{ruledtabular}
\end{table}

\begin{figure*}
  \centering
  \includegraphics[width=0.9\columnwidth]{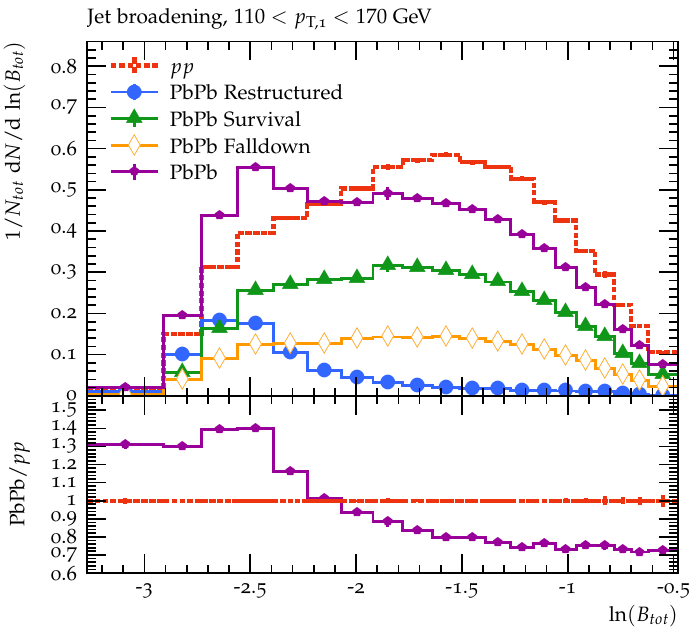}
  \includegraphics[width=0.9\columnwidth]{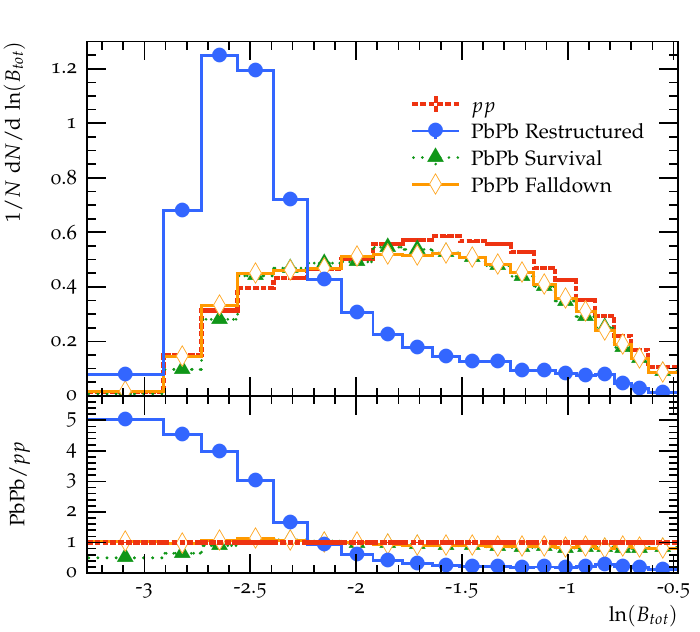}
  \caption{Comparison between the jet broadening \Btot distributions in three categories of quenched events. Left: the distributions of the three types are scaled by the same value, which is the inverse of the total weight of these three distributions. Right: the distributions of the three types are self-normalized.}
  \label{fig:JBDiffSources}
\end{figure*}

We classify the events that pass the cut conditions after energy loss into three categories based on the patterns of the events before jet quenching, and details are given in Table.~\ref{tab:eventcategories}. The tag of ``Survival'' indicates that both the quenched and unquenched versions of those events match the same selection criteria. Some unquenched events satisfy other conditions for selecting events, except that the leading jet $p_{T,1}$ is greater than the maximum limit ($170\GeV$). These events fully satisfy the selection conditions after jet quenching, we record such events as ``Falldown''. In addition to the above two categories, other events are classified as ``Restructured''.
Detailed analysis shows that most ``Restructured'' events come from such unquenched events, that is, the transverse momentum of the leading jet is greater than 110~GeV, and the number of jets greater than 30~GeV is less than three. Specifically, such events account for about 90.2\% of restructured events. For these events, the number of jets with transverse momentum greater than 30~GeV increases after medium modification. For example, when some unquenched events that do not satisfy the event selection conditions pass through the QGP, some jets will split in the medium due to jet-medium interaction, thus the number of jets of these events in the final-state of \pbpb\ increases, and these events then satisfy the selection criteria~\cite{Zhang:2021sua}.

As shown in Fig.~\ref{fig:JBDiffSources}, we plot the distributions of these three categories of quenched events at 5.02~TeV. According to the left plot in Fig.~\ref{fig:JBDiffSources}, we observe that the contributions of these three categories are quite different, with the ``Survival'' part having the most significant contribution at 57.85\%, and the contributions of the ``Falldown'' and the ``Restructured'' are 27.53\% and 14.63\%, respectively. 
We can find that the broad peak of the jet broadening $\lnbtot$ distribution for both the ``Survival'' and the ``Falldown''  of \pbpb\ events, these broad peaks spread from $\lnbtot=-2.6$ to $-1.0$, and the ``Restructured'' part has a sharp peak around $\lnbtot=-2.6$ ($\Btot=0.074$). In fact, the jet broadening distributions of the ``Survival'' and ``Falldown'' parts of \pbpb\ events are not significantly different, nor are they very different from the distribution of \pp\ events. 
It is worth noting that the $\lnbtot$ of the ``Restructured'' part is mainly distributed in regions $-2.9$ to $-2.2$. From Fig.~\ref{fig:JBDiffSources}, we can see that the new contribution from the ``Restructured'' events plays an important role for the enhancement of total jet broadening distributions at small $\lnbtot$.

\begin{figure*} 
  \centering
  \begin{minipage}{0.49\textwidth}
    \centering
    \begin{tikzpicture}
      \node[anchor=south west,inner sep=0] (image) at (0,0) {%
        \includegraphics[width=\textwidth]{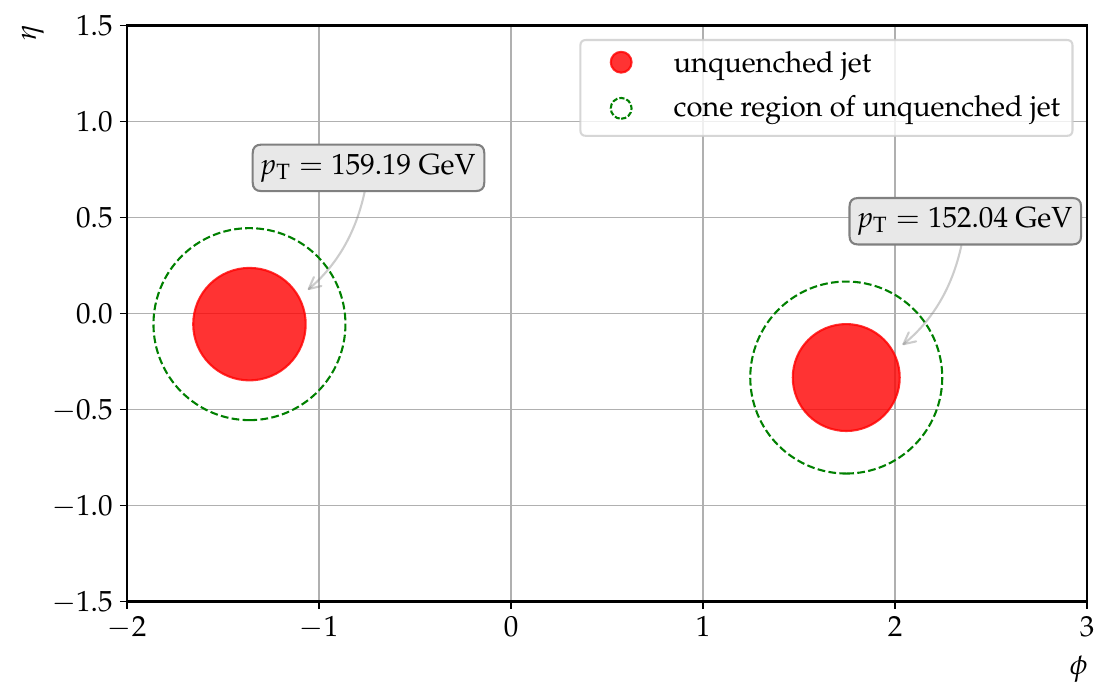}
      };

      \begin{scope}[x={(image.south east)},y={(image.north west)}]
        \node[ultra thick, black] at (0.2, 0.88){(a)};
      \end{scope}
    \end{tikzpicture}
  \end{minipage}
  \begin{minipage}{0.49\textwidth}
    \centering
    \begin{tikzpicture}
      \node[anchor=south west,inner sep=0] (image) at (0,0) {%
        \includegraphics[width=\textwidth]{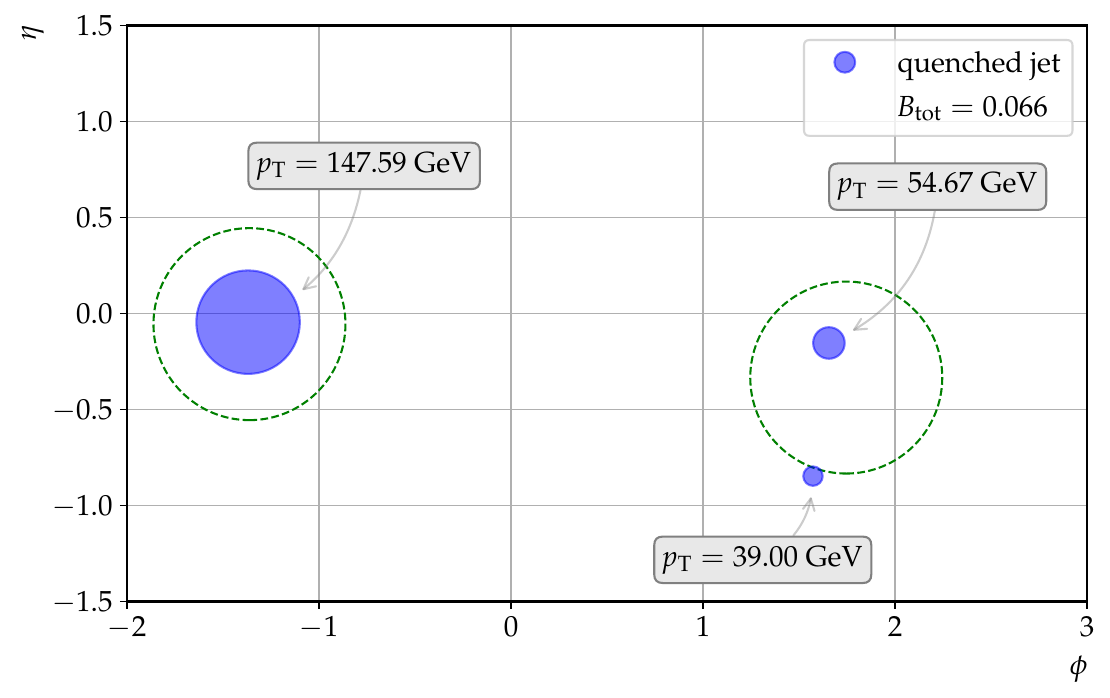}
      };

      \begin{scope}[x={(image.south east)},y={(image.north west)}]
        \node[ultra thick, black] at (0.2, 0.88){(b)};
      \end{scope}
    \end{tikzpicture}
  \end{minipage}
  \caption{Schematic illustration of ``Restructured'' event configuration in the $(\phi, \eta)$-plane. The red and blue disks are unquenched and quenched jets, respectively. The size of the disk represents the relative magnitude of the transverse momentum. The green dashed circle marks the range of the unquenched jet cone.}
  \label{fig:Restructured-Configuration}
\end{figure*}

To further investigate the origins of the ``Restructured'' events, we selected one representative case and plotted the distributions of jets in the $(\eta, \phi)$-plane both before and after jet quenching, as shown in Fig.~\ref{fig:Restructured-Configuration}. The subfigures (a) and (b) in Fig.~\ref{fig:Restructured-Configuration} represent the situations before and after jet quenching, respectively.
Fig.~\ref{fig:Restructured-Configuration} illustrates one possible scenario for the production of ``Restructured'' events, that is, when two hard jets pass through the QGP medium, one jet simply loses energy, while the other jet splits into two smaller jets.
We tried to explore the mechanism of production of ``Restructured'' events by turning off the radiative energy loss mechanism of the LBT model and found that both elastic collision energy loss and radiative energy loss contribute to the production of ``Restructured'' events. We may therefore interpret the ``Restructured'' events as resulting from large-angle elastic scattering of partons within the jet during the in-medium shower, or alternatively, from jet splitting caused by medium-induced gluon radiation. Based on these judgments, the ``Restructured'' events may involve contributions from Moli\`ere scattering.
Moli\`ere scattering in heavy-ion collisions refers to the rare --- though not Gaussianly rare --- large-angle deflection of partons within a jet as they traverse the QGP medium, which is considered a good probe for finding weakly coupled short-distance quark and gluon particles within the strongly coupled liquid QGP~\cite{DEramo:2018eoy}.
We have noticed that large-angle elastic Moli\`ere scattering may contribute to the production of ``Restructured'' events in heavy-ion collisions. However, the contribution of this elastic scattering mechanism is coupled with the contribution from inelastic scattering via medium-induced gluon radiation. Only by considering their combined effects can we achieve the results described earlier.

\section{Summary}
\label{sec:summary}

In this work, we present the first theoretical result of the medium modifications of jet broadening distribution due to the jet quenching effect in heavy-ion collisions at large momentum transfer.
The \pp\ baseline is provided by \POWHEG$+$\PYTHIAe, where \POWHEG\ is responsible for generating NLO precision hard process events and \PYTHIAe\ executes procedures such as parton shower. We use the linear Boltzmann transport (LBT) model for simulating parton energy loss to perform the evolution of jets in the medium.

The jet broadening observable $\Btot$, which belongs to event-shape variables, can be used to characterize the broadening degree of multijet energy flow. The decrease in this variable indicates that the jets' distribution within the event becomes narrower.
We calculate the medium modification factor as a function of jet broadening variable $\Btot$. Compared to \pp\ references, the results show an enhancement at the small \Btot region but a suppression at large \Btot in \pbpb\ collisions, which implies the medium modification leads to a narrower distribution of jet broadening relative to \pp\ references.

We further explore the possible underlying reasons for the medium modification of jet broadening distributions. We demonstrate that jet broadening distributions for three-jet events and for $\njet \geqslant 4$ events are pretty different in both \pp\ and \pbpb\ collisions at \snn{5.02}{TeV}. Due to parton energy loss in the QGP,  the relative production fraction of $\njet \geqslant 4$ events may be reduced. This jet number reduction effect then results in the decreasing of jet broadening distributions in the large \Btot region as well as an increasement  in the low \Btot region, in high-energy nuclear collisions.
Furthermore, jet-medium interactions may give rise to a new contribution from the ``Restructured'' events, which further increases jet broadening distributions  at small  \Btot in \pbpb\ collisions.
The preliminary study shows that the jets of most ``Restructured'' events will occur splitting when they pass through the QGP medium, leading to an increase in the number of jets in the event.

The narrowing of jets reported in this work is the behavior of multijet's energy flow geometric patterns within the event after nuclear modification.
We noticed an interesting phenomenon: measurements of many jet substructure observables show that the energy flow inside the jet also exhibits a narrowing behavior after jet quenching. These observables include the groomed jet radius~\cite{Casalderrey-Solana:2019ubu,Ringer:2019rfk,ALargeIonColliderExperiment:2021mqf,Caucal:2021cfb,Cunqueiro:2021wls,Wang:2022yrp}, the momentum splitting fraction~\cite{ALICE:2019ykw,Casalderrey-Solana:2019ubu,Caucal:2019uvr}, the jet mass~\cite{Casalderrey-Solana:2019ubu}, the nuclear modification factor correlated with splitting angular distance~\cite{ATLAS:2022vii}, or jet shape observables like the girth, the momentum dispersion~\cite{ALICE:2018dxf,Cunqueiro:2021wls}, etc.
The physical mechanisms responsible for the narrowing of jets observed in such jet substructure observables are not always the same. Several different mechanisms have been introduced to explain, for instance, jet energy loss can lead to an increased fraction of quark-initiated jets, which are typically narrower~\cite{Ringer:2019rfk,Pablos:2022mrx,ALargeIonColliderExperiment:2021mqf}; alternatively, narrower jets usually lose less energy in the medium and thus have a higher chance of surviving~\cite{Caucal:2019uvr,Casalderrey-Solana:2019ubu,Cunqueiro:2021wls}.
Unlike such jet substructure observables, the narrowing of jet broadening distribution is primarily due to the reduction of multijet ($\njet\geqslant 4$) events caused by energy loss, and it is precisely these multijet events that represent a wider jet broadening distribution.

\paragraph*{Acknowledgments:} We are grateful to S Y Chen and Q Zhang for providing helpful
suggestions. We acknowledge support for our work from following open-source projects, such as \POWHEG~\cite{Nason:2004rx,Frixione:2007vw,Alioli:2010xd,Alioli:2010xa}, \PYTHIAe~\cite{Sjostrand:2014zea}, \FASTJET~\cite{Cacciari:2011ma}, \textsc{HepMC3}~\cite{Buckley:2019xhk}, \textsc{Rivet}~\cite{Bierlich:2019rhm}, and \textsc{Matplotlib}~\cite{Hunter:2007}. 
This work is supported by the Guangdong Major Project of Basic and Applied Basic Research No. 2020B030103008, and Natural Science Foundation of China with Project Nos. 11935007 and 12035007.

\appendix

\section{The symmetric multijet limit}
\label{sec:appendix}

The main text mentioned the maximum values of transverse thrust and jet broadening observables with given restrictions (ignore the $p_z$ of the final state jets). Here we provide detailed proofs, following the discussions in Ref.~\cite{Banfi:2010xy}.
For a symmetrical planar transverse event with jet numbers $N$ of equivalent momenta, the component jets can be expressed in the $(p_T, \eta, \phi)$-coordinate system as $p_i=(p_T, 0, \frac{2\pi}{N}i)$ for $i=1,2,\cdots,N$.

\subsection{Transverse thrust}

For a symmetrical multijet event with $N$ jets of equivalent momentum, the specific direction of the transverse thrust axis is unimportant, but what important is the angle between it and the nearest jet beside it, which we denote here as $\Delta\phi$.
So the transverse thrust $\TThrust$ can be expressed as
\begin{eqnarray}
  \TThrust &\equiv& 1 - \max_{\hat{n}_T} \frac{\sum_i \left|%
    \vec{p}_{T,i}\cdot \hat{n}_T\right|
  }{\sum_i p_{T,i}} \nonumber \\
  &=& 1 - \frac{p_T\sum_{i=1}^{N}\left|\cos\left(\frac{2\pi i}{N}-\Delta\phi\right)\right|}{N\cdot p_T},
\end{eqnarray}
where $N$ is the number of jets and $N=2\pi/\phi$. We can get the $\TThrust$ value for perfectly circular planar events ($N\to\infty, \phi\to 0$),
\begin{eqnarray}
  \left.\TThrust\right|_{\max}
  &=& \lim_{\phi\to 0} 1 - \frac{\sum_{i=1}^{N}\left|\cos\left(i\phi-\Delta\phi\right)\right|}{2\pi/\phi} \nonumber\\
  &=& 1-\frac{2}{\pi},
\end{eqnarray}
notice here that when $\phi\to 0$, must also have $\Delta\phi\to 0$.

\subsection{Jet broadening}

One important aspect is to separate the event into the upper and lower hemispheres when calculating the jet broadening. As the interface between these hemispheres is perpendicular to the transverse thrust axis, it is crucial to ensure that no jets are present on this interface in order to maximize the sum of the absolute values of all jets' projections on the transverse thrust axis.

When $N$ is an even number, we choose the center of the first and $N$-th jet as the interface, from which we can determine that $\phi_U$ and $\phi_L$ as $\pi/N+\pi/2$ and $\pi/N-\pi/2$, respectively. Then, we can get
\begin{eqnarray}
  B_U &=& \frac{1}{2Np_T}\sum_{i\in\mathcal{C}_U}p_T\sqrt{\left(\frac{2\pi i}{N}-\frac{\pi}{N}-\frac{\pi}{2}\right)^2}\nonumber\\
  &=& \frac{1}{4N^2}\sum_{i=1}^{N/2}\left|4\pi i - 2\pi - N\pi\right|.
\end{eqnarray}
Since the lower side and the upper side are completely symmetrical, and $\phi_U$ and $\phi_L$ are on the same line, $B_L$ and $B_U$ are exactly the same, we have
\begin{eqnarray}
  \Btot &=& 2B_U = \frac{1}{2N^2}\sum_{i=1}^{N/2}\left|4\pi i - 2\pi - N\pi\right| \nonumber\\
  &=& \frac{\pi}{N^2}\sum_{i=1}^{\lfloor N/4\rfloor}\left(2+N-4i\right),\nonumber\\
  &=& \frac{\pi}{N^2}\left(N-2\left\lfloor\frac{N}{4}\right\rfloor\right)\left\lfloor\frac{N}{4}\right\rfloor.
  \label{app:eq:BtotEven}
\end{eqnarray}

When $N$ is an odd number, the unit direction vector of any jet can be chosen as transverse thrust axis $\hat{n}_T$,
\begin{widetext}
  \begin{eqnarray}
    \Btot = B_U + B_L
    &=& \frac{1}{2Np_T}\sum_{i\in\mathcal{C}_U}p_T\sqrt{\left(\frac{2\pi i}{N}\right)^2}
    + \frac{1}{2Np_T}\sum_{i\in\mathcal{C}_L}p_T\sqrt{\left(\frac{2\pi i}{N} - \frac{\pi}{N}\right)^2}\nonumber\\
    &=& \frac{2\pi}{N^2}\sum_{i=1}^{\lfloor (N-1)/4 \rfloor}i + \frac{1}{N^2}\sum_{i=1}^{\lceil (N-1)/4\rceil}(2\pi i - \pi) \nonumber\\
    &=& \frac{\pi}{N^2} \left(\left\lceil \frac{N-1}{4}\right\rceil ^2+\left\lfloor \frac{N-1}{4}\right\rfloor ^2+\left\lfloor \frac{N-1}{4}\right\rfloor \right).
    \label{app:eq:BtotOdd}
  \end{eqnarray}
\end{widetext}
For perfectly circular planar events ($N\to\infty$), both Eq.~\eqref{app:eq:BtotEven} and Eq.~\eqref{app:eq:BtotOdd} tend to $\pi/8$.

\bibliographystyle{apsrev4-1}
\bibliography{refs}

\end{document}